\documentclass[prb,twocolumn,showpacs,superscriptaddress,amsmath,amssymb]{revtex4}
\usepackage{mathrsfs}

\usepackage[colorlinks,linkcolor=blue,citecolor=blue]{hyperref}
\usepackage{graphicx}
\usepackage{amssymb}
\usepackage{dcolumn}
\usepackage{bm}
\usepackage{color}

\begin{document}


\title{Quantized multi-terminal resistances in ferromagnetic graphene \emph{p-n} junctions}

\author{Lei Xu}
\affiliation{National Laboratory of Solid State Microstructures and
Department of Physics, Nanjing University, Nanjing 210093, China}

\author{Jin An}
 \email{anjin@nju.edu.cn}
\affiliation{National Laboratory of Solid State Microstructures and
Department of Physics, Nanjing University, Nanjing 210093, China}

\author{Chang-De Gong}
\email{cdgongsc@nju.edu.cn}
\affiliation{Center for Statistical and Theoretical Condensed Matter
Physics, and Department of Physics, Zhejiang Normal University,
Jinhua 321004, China}

\affiliation{National Laboratory of Solid State Microstructures and
Department of Physics, Nanjing University, Nanjing 210093, China}

\date{\today}

\begin{abstract}

The quantum Hall and longitudinal resistances in multi-terminal ferromagnetic graphene \emph{p-n} junctions under a perpendicular magnetic field are investigated. In the Hall measurements, the transverse contacts are assumed to be located at \emph{p-n} interface to avoid the mixing of edge states at the interface and the resulting quantized resistances are then topologically protected. According to the charge carrier type, the resistances in four-terminal \emph{p-n} junction can be naturally divided into nine different regimes. The symmetric Hall and longitudinal resistances are observed, with lots of new robust quantum plateaus revealed due to the competition between spin splitting and local potentials.

\end{abstract}


\maketitle

Graphene, a single-layer hexagonal lattice of carbon atoms, has recently attracted extensive interest.\cite{Castro Neto2009} For neutral
graphene, Fermi energy is located at the two inequivalent Dirac points, around which graphene has a unique band structure with a linear dispersion, giving it many peculiar properties such as the half-integer quantum Hall
(QH) effect.\cite{Castro Neto2009,Novoselov2005,Zhang2005,Geim2007} The corresponding QH edge state in graphene has been proved to be
chiral\cite{Novoselov2005,Zhang2005} with the chirality determined by the carrier type and the direction of the magnetic field. Thus, to manipulate the edge-state transmission in graphene, one may vary both the carrier type and concentration of graphene by the gate voltages\cite{Huard2007,Williams2007} or by chemically doping the underlying
substrate.\cite{Ohta2006}

\begin{figure}
\scalebox{0.8}[0.75]{\includegraphics[138,488][429,753]{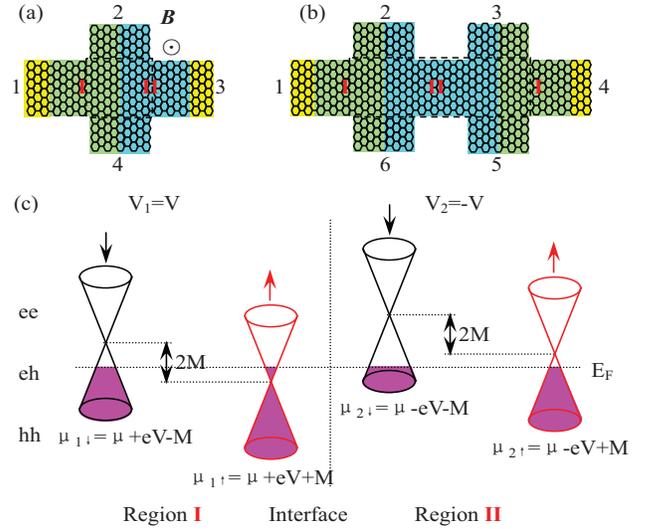}}  \caption{\label{fig1}(Color online) The top panels (a) and (b) are the schematic for the four- and six-terminal FGPNJ devices. The whole devices, made of the ferromagnetic graphene, are divided into region \textbf{I} and region \textbf{II}) by the interfaces of \emph{p-n} junctions. The bottom panel (c) shows schematically for the 4-terminal situation the charge carrier types in both regions with potential difference $2V$, where $\mu_{1\sigma}$ and $\mu_{2\sigma}$ are the effective chemical potentials for $\sigma$ spin component and $E_F$ the Fermi energy. The symbol ``e (h)'' represents that the carriers are electrons (holes).}
\end{figure}

On the other hand, a graphene \emph{p-n} junction, of which carrier density in two adjacent regions can be individually controlled by a pair of
gate voltages above and below it was reported.\cite{Williams2007} Many new and exciting phenomena are predicted, such as Klein tunneling,\cite{Katsnelson2006,Stander2009} particle collimation,\cite{Cheianov2006} quasibound states,\cite{Silvestrov2007,Nguyen2009} and Veselago lensing.\cite{Cheianov2007} In
addition, the electron transport through the graphene \emph{p-n} or \emph{p-n-p} junctions has stimulated many subsequent researches.\cite{Tworzydlo2007,Akhmerov2008,Cresti2008,Fistul2007,Abanin2007,Ozyilmaz2007,Xing2010,Ki2009,Lohmann2009,Ki2010,Chen2010,Velasco2010} In the two-terminal measurements, QH plateaus show half-integer values in
the case of unipolar junctions but fractional values for bipolar junctions.\cite{Williams2007} The appearance of the fractional plateaus is well
explained by assuming full mixing of the electronlike and holelike edge states at \emph{p-n} interface.\cite{Abanin2007} The experimentally observed conductance plateaus in two-terminal measurements are often distorted, because the conductance generally depends on the sample geometry and the interface inhomogeneities.\cite{Huard2008,Abanin2008,Williams2009,Low2009-1,Low2009-2} Thus, it makes sense to seek the measurements in which the QH plateaus are geometry-independent and do not rely on the edge-state equilibration at \emph{p-n} interface. Different from the previous transport measurements,\cite{Ki2009,Lohmann2009,Ki2010,Chen2010,Velasco2010} here we propose multi-terminal measurements in
which the transverse contacts are located at \emph{p-n} interfaces [see Figs.~\ref{fig1}(a) and \ref{fig1}(b)]. In these proposed measurements, additional interface conduction channels appear between the transverse terminals\cite{Williams2011} and the current-carrying edge states directly propagate along interface without the mode mixing effect which is unavoidable in the measurement of previous studies. Since electron transmission between any two contacts is dominated by an integer number of edge states, the measured resistance plateaus in this situation should be topologically protected, which is expected to have a potential application in spintronic devices.

When the Fermi energy of a ferromagnetic graphene\cite{Haugen2008,Linder2008} is set between the spin-up and spin-down Dirac points [i.e., Fermi energy lies in the spin gap $2M$ in Fig.~\ref{fig1}(c)], the carriers consist of spin-up electrons and spin-down holes, each type of which is completely spin-polarized.\cite{Sun2010} Therefore, one can expect more interesting phenomena of quantum transport in the ferromagnetic graphene \emph{p-n} junctions (FGPNJ). In this paper, using the tight-binding model, we carry out a theoretical study on the multi-terminal resistances in FGPNJ in the proposed situation. Here we focus on the four-terminal and also discuss six-terminal Hall measurements, and we reveal some new fractional plateaus of QH and longitudinal resistances across the junctions. These new fractional plateaus are predicted to be robust against weak disorder, which is confirmed by a subsequent numerical calculation.

The Hamiltonian for the four- or six-terminal FGPNJ in the tight-binding representation is given by,
\begin{equation}
H=\sum_{i,\sigma}(\epsilon_i-\mu)\hat{c}_{i,\sigma}^\dag\hat{c}_{i,\sigma}-
\sum_{<ij>,\sigma}(te^{i\phi_{ij}}\hat{c}_{i,\sigma}^\dag\hat{c}_{j,\sigma}+\mathrm{H.c.}), \label{eq:1}
\end{equation}
where $\hat{c}_{i,\sigma}^\dag$ and $\hat{c}_{i,\sigma}$ are the creation and annihilation operators
with spin $\sigma=\pm1$ at site $i$. $\epsilon_i$ is the energy of Dirac points (i.e. the on-site energy),
and $\epsilon_i=-eV-\sigma M$ for region \textbf{I} and $\epsilon_i=eV-\sigma M$ for region \textbf{II} with
$\pm V$ the electric potentials which can be controlled by the gate voltages in the experiment. $\mu$ is the chemical potential. $M$ is the ferromagnetic exchange split.\cite{Haugen2008} $t$ is the nearest-neighbor hopping integral.
In the presence of a perpendicular magnetic field $\mathbf{B}$, a Berry phase factor $\phi_{ij}=\int_i^j\mathbf{A}\cdot d\mathbf{l}/\phi_0$ is added to the hopping integral with $\mathbf{A}$ the magnetic gauge potential and $\phi_0=h/e$ the flux quantum.

The Landau levels (LLs) of graphene in a uniform magnetic field $B$, is described by $\varepsilon_n=\pm\varepsilon\sqrt{|n|}$ with $\varepsilon=\upsilon_F\sqrt{2e\hbar B}$ the energy of the first LL, $v_F$ the Fermi velocity, and $n=0,\pm1,\pm2,...$ the LL index.
For FGPNJ, the effective chemical potentials which are measured from the Dirac points can be expressed as $\mu_{1\sigma}=\mu+eV+\sigma M$ for region \textbf{I} and
$\mu_{2\sigma}=\mu-eV+\sigma M$ for region \textbf{II}. With $\mu_{i\sigma}$ measured in unit of $\varepsilon$, the LL filling factors can be given by
\begin{eqnarray}
&&\nu_{i\sigma}=2[\mathbf{sgn}(\mu_{i\sigma})(\mu_{i\sigma})^2]+1,\label{eq:2}
\end{eqnarray}
where $i=1,2$, and $[x]$ is the maximal integer no more than $x$.

\begin{figure}
\scalebox{0.75}[0.75]{\includegraphics[150,542][425,754]{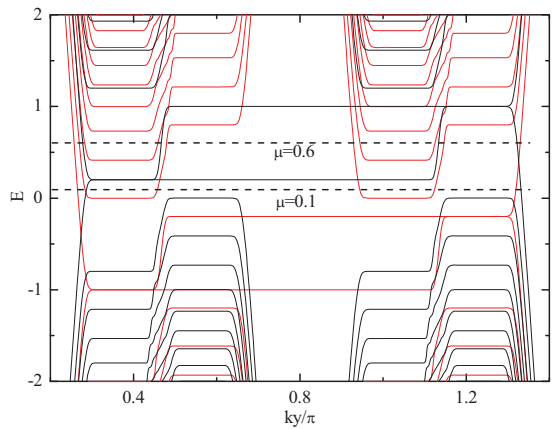}}  \caption{\label{fig2}(Color online) Electron energy spectrum of a two-terminal FGPNJ (see
Fig.~\ref{fig1}(a) without the longitudinal terminals 1and 3) in the presence of a perpendicular magnetic field $B$, where the Landau gauge $\mathbf{A}=(0,Bx,0)$ is adopted. The red and black lines denote bands of spin-up
and spin-down states. The dashed lines represent the positions of the chemical potentials. Energy and chemical potentials are measured by
$\varepsilon=\upsilon_F\sqrt{2e\hbar B}$ with $B=10$~T, $M=0.6$ ($\sim 60$~meV) and $eV=0.4$ ($\sim 40$~meV).}
\end{figure}

In Fig.~\ref{fig2}, we give the energy spectrum of a two-terminal FGPNJ as an example to find the redistribution behavior of filling factors owing to spin splitting and nonuniform local potentials. The local potential we consider here is assumed to change abruptly at the interface because the interface width can be very small.\cite{Abanin2008,Williams2009} We find that, for each spin component, there are two groups of LLs corresponding to region \textbf{I} and region \textbf{II}. By counting the number of electronlike (holelike) LLs below (above) the chemical potentials (the dash lines in Fig.~\ref{fig2}), it is easily to determine the LL filling factors. In Fig.~\ref{fig2}, for $\mu=0.1$, the filling factors are $(\nu_{1\uparrow},\nu_{1\downarrow})=(3,-1)$ and
$(\nu_{2\uparrow},\nu_{2\downarrow})=(1,-1)$, and for $\mu=0.6$, $(\nu_{1\uparrow},\nu_{1\downarrow})=(5,1)$ and
$(\nu_{2\uparrow},\nu_{2\downarrow})=(1,-1)$. These estimation are exactly consistent with that calculated directly by Eq.~(\ref{eq:2}). Compared with the filling factors in pristine graphene $(\nu_\uparrow,
\nu_\downarrow)=(2n+1,2n+1)$ with $n$ the LL index, the sequence of the filling factors are completely different, from which the new QH plateaus may be expected to arise. In the following, we first give the expression for the quantized resistances and then we make numerical calculations to confirm the predictions.

\begin{figure}
\scalebox{0.75}[0.75]{\includegraphics[135,298][429,731]{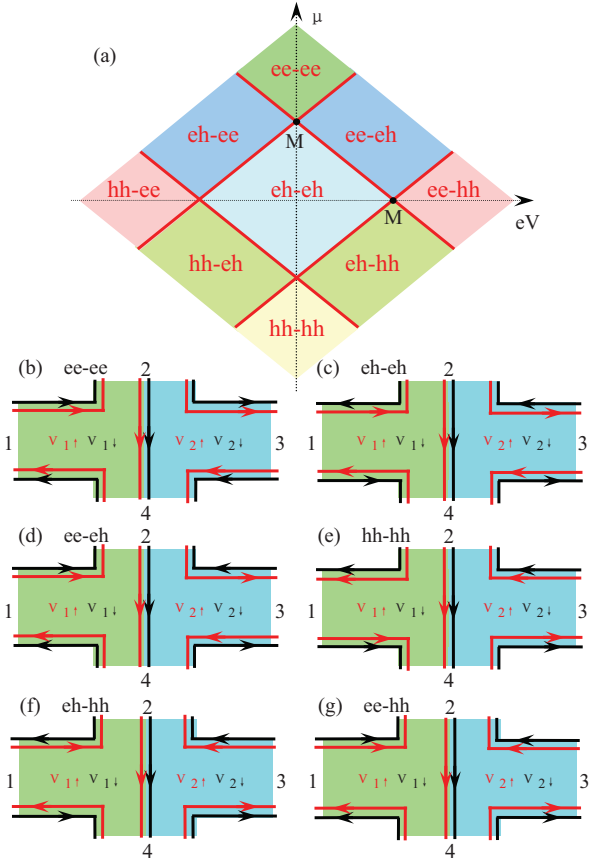}}  \caption{\label{fig3}(Color online) (a) Schematic diagram of the resistance depending on $\mu$ and $eV$ in FGPNJ. The resistance is distinguished into nine regimes which are separated by red solid lines and marked by different color. (b)-(g) Diagrams for edge states propagation in the corresponding regimes for $V>0$, where $\nu_{1\sigma}$ and $\nu_{2\sigma}$ are the filling factors in region \textbf{I} and region \textbf{II}. The red(black) lines with arrows represent spin-up(spin-down) edge states.}
\end{figure}

Firstly we consider the four-terminal Hall device [see Fig.~\ref{fig1}(a)]. To investigate the QH resistance $R_{13,24}$ and longitudinal resistance $R_{13,13}$, which are defined by $R_{13,24}=(V_2-V_4)/I_1$ and $R_{13,13}=(V_1-V_3)/I_1$, we make voltages of terminal 2 and 4 floating, and apply a small bias voltage $V_0$ between terminal 1 and 3, leading to a steady current $I_1$ flowing between the two terminals. From the Landauer-B\"{u}ttiker formula,\cite{Datta1995} the electric current injected from terminal $p$ is given by
\begin{equation}
I_p=\frac{e^2}{h}\sum_q T_{pq}(V_p-V_q). \label{eq:4}
\end{equation}
Here $T_{pq}$ is the transmission coefficient from terminal $q$ to terminal $p$, and $V_q$ the voltage of terminal $q$. In the QH regime of the situation shown in Fig.~\ref{fig1}(a), except $T_{13}$=$T_{31}$=0, $T_{pq}$ between any other two terminals is an integer which depends on the four filling factors $\nu_{i\sigma}$ (i=1,2, $\sigma$=$\uparrow$,$\downarrow$), since electron transmission between any two terminals is dominated by edge-state propagation. Therefore, according to the positions of the effective chemical potentials, one can classify the system into nine different regimes [see Fig.~\ref{fig3}(a)], which are divided by four straight lines $\mu_{i\sigma}=0$ ($i=1,2$, $\sigma$=$\uparrow$,$\downarrow$). In FGPNJ, we have $\nu_{1\sigma}(\mu,\pm V,M)=\nu_{2\sigma}(\mu,\mp V,M)$, which is equivalent to changing \emph{p-n} junction to \emph{n-p} junction without any new physics in it. Thus, we only need to discuss the six regimes for $V>0$ shown in Figs.~\ref{fig3}(b)-\ref{fig3}(g).

For ee-ee regime, where the first (second) ``e'' before ``-'' indicates the carrier type in Region \textbf{I} in spin-up (spin-down) component is electron-like while the symbol ``ee'' after ``-'' has a similar meaning, the nonzero transmission coefficients are $T_{21}=T_{14}=|\overline{\nu_1}|$ and $T_{32}=T_{43}=|\overline{\nu_2}|$ with
$|\overline{\nu_1}|=|\nu_{1\uparrow}|+|\nu_{1\downarrow}|$, $|\overline{\nu_2}|=|\nu_{2\uparrow}|+|\nu_{2\downarrow}|$. The currents $I_1=-I_3=|\overline{\nu_2}|V_0e^2/h$, and the voltage differences $V_1-V_3=V_0$ and
$V_2-V_4=(|\overline{\nu_2}|/|\overline{\nu_1}|)V_0$, the QH resistance $R_{13,24}$ and the longitudinal resistance $R_{13,13}$ are calculated ($|\nu_{1\uparrow}|\geq|\nu_{2\uparrow}|$, $|\nu_{1\downarrow}|\geq|\nu_{2\downarrow}|$),
\begin{eqnarray}
&&R_{13,24}=\frac{h}{e^2}\frac{1}{|\overline{\nu_1}|},\label{eq:5}\\
&&R_{13,13}=\frac{h}{e^2}\frac{1}{|\overline{\nu_2}|}.\label{eq:6}
\end{eqnarray}

Similarly, one can obtain the resistances $R_{13,24}$ and $R_{13,13}$ in the other five regimes
and detailed results are summarized as follow:

(a) eh-eh regime ($|\nu_{1\uparrow}|\geq|\nu_{2\uparrow}|$, $|\nu_{1\downarrow}|\leq|\nu_{2\downarrow}|$):
\begin{eqnarray}
&&R_{13,24}=\frac{h}{e^2}\frac{|\nu_{1\uparrow}||\nu_{2\uparrow}|-|\nu_{1\downarrow}||\nu_{2\downarrow}|}
{|\nu_{2\downarrow}|^2|\overline{\nu_1}|+|\nu_{1\uparrow}|^2|\overline{\nu_2}|},\label{eq:7}\\
&&R_{13,13}=\frac{h}{e^2}\frac{(|\nu_{1\uparrow}|+|\nu_{2\downarrow}|)^2}
{|\nu_{2\downarrow}|^2|\overline{\nu_1}|+|\nu_{1\uparrow}|^2|\overline{\nu_2}|};\label{eq:8}
\end{eqnarray}
(b) ee-eh regime ($|\nu_{1\uparrow}|\geq|\nu_{2\uparrow}|$):
\begin{eqnarray}
&&R_{13,24}=\frac{h}{e^2}\frac{|\nu_{2\uparrow}|}
{|\nu_{2\downarrow}|^2+|\overline{\nu_1}||\overline{\nu_2}|},\label{eq:9}\\
&&R_{13,13}=\frac{h}{e^2}\frac{(|\overline{\nu_1}|+|\nu_{2\downarrow}|)^2}
{(|\nu_{2\downarrow}|^2+|\overline{\nu_1}||\overline{\nu_2}|)|\overline{\nu_1}|};\label{eq:10}
\end{eqnarray}
(c) hh-hh regime ($|\nu_{1\uparrow}|\leq|\nu_{2\uparrow}|$, $|\nu_{1\downarrow}|\leq|\nu_{2\downarrow}|$):
\begin{eqnarray}
&&R_{13,24}=-\frac{h}{e^2}\frac{1}{|\overline{\nu_2}|},\label{eq:11}\\
&&R_{13,13}=\frac{h}{e^2}\frac{1}{|\overline{\nu_1}|};\label{eq:12}
\end{eqnarray}
(d) eh-hh regime ($|\nu_{1\downarrow}|\leq|\nu_{2\downarrow}|$):
\begin{eqnarray}
&&R_{13,24}=-\frac{h}{e^2}\frac{|\nu_{1\downarrow}|}
{|\nu_{1\uparrow}|^2+|\overline{\nu_1}||\overline{\nu_2}|},\label{eq:13}\\
&&R_{13,13}=\frac{h}{e^2}\frac{(|\nu_{1\uparrow}|+|\overline{\nu_2}|)^2}
{(|\nu_{1\uparrow}|^2+|\overline{\nu_1}||\overline{\nu_2}|)|\overline{\nu_2}|};\label{eq:14}
\end{eqnarray}
(e) ee-hh regime:
\begin{eqnarray}
&&R_{13,24}=0,\label{eq:15}\\
&&R_{13,13}=\frac{h}{e^2}(\frac{1}{|\overline{\nu_1}|}+\frac{1}{|\overline{\nu_2}|}).\label{eq:16}
\end{eqnarray}

\begin{figure*}
\scalebox{1}[1]{\includegraphics[72,260][560,639]{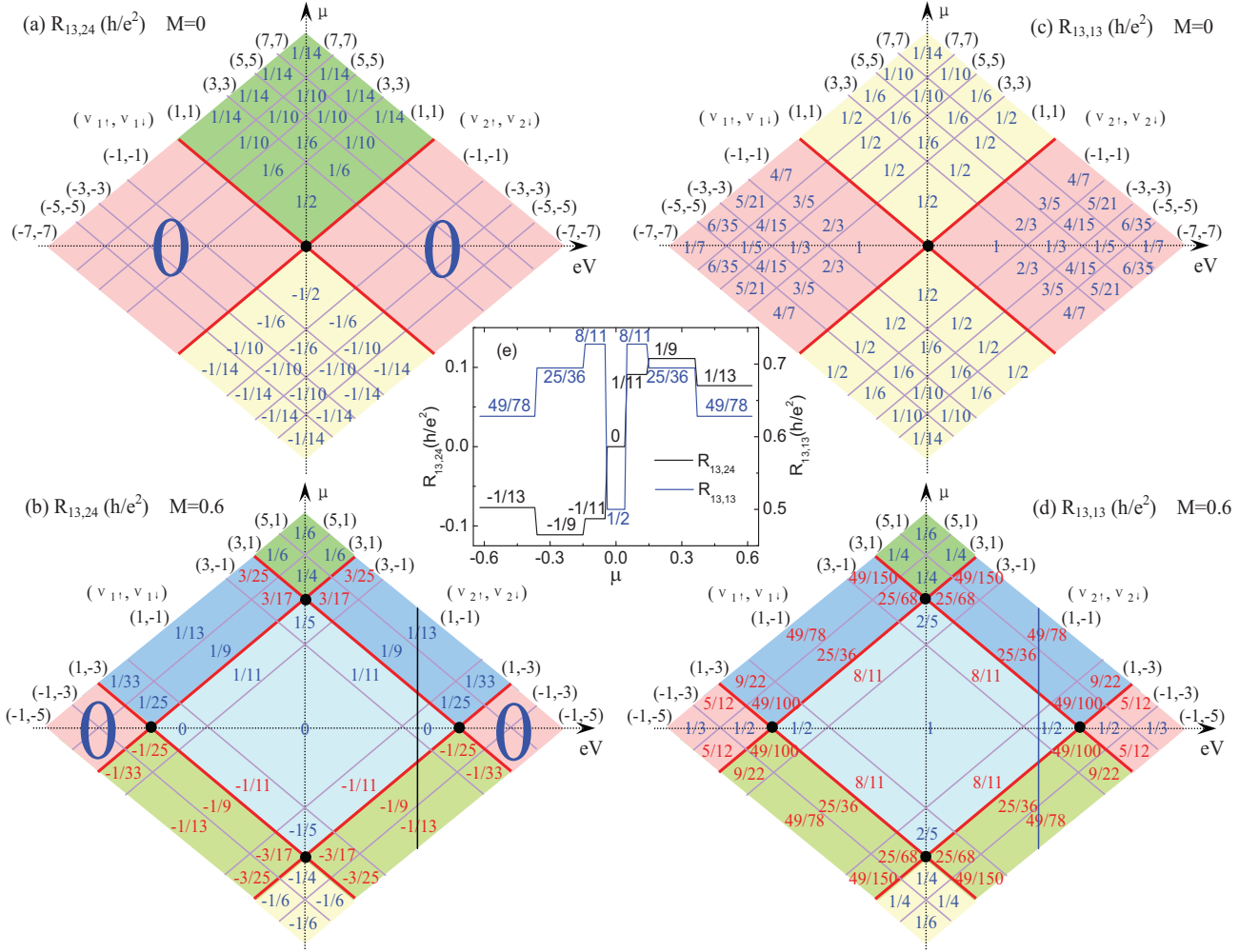}} \caption{\label{fig4}(Color online) Simulated color maps of the theoretical QH resistance $R_{13,24}$ [(a) and (b)] and longitudinal resistance $R_{13,13}$ [(c) and (d)] expected from the mechanisms shown in Fig.~\ref{fig3} for different filling factors $(\nu_{1\uparrow},\nu_{1\downarrow})$ and $(\nu_{2\uparrow},\nu_{2\downarrow})$ in region \textbf{I} and region \textbf{II} with parameters $M=0$ and $0.6$. The coordinates of the points ``$\bullet$'' are $(\pm M,0)$ and $(0,\pm M)$. (e) The one-dimensional slices shown in (b) (black solid) and (d) (blue solid) are plotted to show the dependence of $R_{13,24}$ and $R_{13,13}$ on $\mu$ for a fixed $eV=0.45$.}
\end{figure*}

Based on the above equations, we calculate the QH and longitudinal resistances and plot the results for $M=0$ and $0.6$ in Fig.~\ref{fig4}. It displays several blocks of different sequence of fractional quantized resistances. In ee-hh regime, the QH resistances are always zero regardless of $M$, $\mu$ and $V$, but the longitudinal resistance are the summation of the contributions of region \textbf{I} and region \textbf{II}. Apart form the trivial cases of ee-ee, hh-hh, and ee-hh, the other nontrivial cases give new resistance plateaus. Similar to Ref.~\onlinecite{Chen2010}, both $R_{13,24}$ and $R_{13,13}$ have a mirror symmetry with respect to the $\mu$-axis,
\begin{eqnarray*}
R_{13,24}(\nu_{1\uparrow},\nu_{1\downarrow};\nu_{2\uparrow},\nu_{2\downarrow})=
R_{13,24}(\nu_{2\uparrow},\nu_{2\downarrow};\nu_{1\uparrow},\nu_{1\downarrow}),\\
R_{13,13}(\nu_{1\uparrow},\nu_{1\downarrow};\nu_{2\uparrow},\nu_{2\downarrow})=
R_{13,13}(\nu_{2\uparrow},\nu_{2\downarrow};\nu_{1\uparrow},\nu_{1\downarrow}),
\end{eqnarray*}
owing to the equivalence between \emph{p-n} and \emph{n-p} junctions, i.e., the equivalence between Hamiltonians $H(B,\mu,-V)$ and $H(B,\mu,V)$. Furthermore, $R_{13,24}$ are antisymmetric whereas $R_{13,13}$ are symmetric with respect to the $V$-axis,
\begin{eqnarray*}
R_{13,24}(\nu_{1\uparrow},\nu_{1\downarrow}&;&\nu_{2\uparrow},\nu_{2\downarrow})=\\
&&-R_{13,24}(-\nu_{2\downarrow},-\nu_{2\uparrow};-\nu_{1\downarrow},-\nu_{1\uparrow}),\\
R_{13,13}(\nu_{1\uparrow},\nu_{1\downarrow}&;&\nu_{2\uparrow},\nu_{2\downarrow})=\\
&&R_{13,13}(-\nu_{2\downarrow},-\nu_{2\uparrow};-\nu_{1\downarrow},-\nu_{1\uparrow}),
\end{eqnarray*}
as a result of a combination symmetry of four-terminal FGPNJ. This combination symmetry means that $\mathcal{R}H(B,-\mu,V)\mathcal{R}^{-1}$ is equivalent to $H(B,\mu,V)$, except the exchange of terminals 1 and 3, where $\mathcal{R}=\mathcal{P}\mathcal{T}$ with $\mathcal{P}$ the particle-hole transformation [$c_{i\sigma}\rightarrow \eta c_{i\sigma}^\dag$ with $\eta=1$ ($-1$) if $i$ belongs to $A$ ($B$) sublattice] and $\mathcal{T}$ the time-reversal transformation.

For $M=0$ [see Figs.~\ref{fig4}(a) and \ref{fig4}(c), there is no spin splitting, $\nu_{1\uparrow}=\nu_{1\downarrow}$ and $\nu_{2\uparrow}=\nu_{2\downarrow}$], there are only four different carrier-type regimes (ee-ee, hh-hh, hh-ee and ee-hh regimes) and the results are in exact agreement with recent experiment\cite{Williams2007} and theory.\cite{Abanin2007} This indicates that in this situation the presence of the transverse contacts not only does not change the resistances, but also make the resistance plateaus topologically stable. For $M\neq 0$, many new fractional values [see the values in red color in Figs.~\ref{fig4}(b) and \ref{fig4}(d)] are induced.\footnote{The new fractional values exclude the values in the case with $M=0$ and
$V\neq 0$ [see Figs.~\ref{fig4}(a) and \ref{fig4}(c)], and in the case with $M\neq 0$ and $V=0$ which has been discussed in
Ref.~\onlinecite{Sun2010} where the resistances are $R_{13,24}=(|\nu_\uparrow|-|\nu_\downarrow|)/(|\nu_\uparrow|^2+|\nu_\downarrow|^2)$ and
$R_{13,13}=(|\nu_\uparrow|+|\nu_\downarrow|)/(|\nu_\uparrow|^2+|\nu_\downarrow|^2)$.} These new fractional values of QH and longitudinal resistances are caused by spin splitting together with local potentials. More details can be seen in Fig.~\ref{fig4}(e), which is extracted from Figs.~\ref{fig4}(b) (black curve) and \ref{fig4}(d) (blue curve) for fixed $eV=0.45$. It shows that, upon increasing the chemical potential gradually by tuning the gate voltage, the QH resistance undergoes the plateau values (in the unit of $h/e^2$) $\pm1/13$, $\pm1/9$, $\pm1/11$ and $0$, where the values $-1/13$, $-1/11$ and $-1/9$ are newly revealed, while the longitudinal resistance exhibits the plateau values $49/78$, $25/36$, $8/11$ and $1/2$ with the newly induced values $49/78$, $25/36$ and $8/11$. We remark that since energy is measured in unit of $\varepsilon$, the conclusion obtained in Fig.~\ref{fig4} is actually universal, i.e., it is independent of the magnitude of the magnetic field, so long as the system is within the QH regime and the magnetic length $l_B$ is much less than size of the device. Different from the situation discussed in previous studies,\cite{Williams2007,Abanin2007,Ozyilmaz2007,Ki2009,Lohmann2009,Ki2010} where the edge states with different filling factors are equilibrated at \emph{p-n} interface, the quantized resistances obtained here in the proposed situation are topological stable, since the transverse contacts are just located at \emph{p-n} interface and there is no mixing of edge states at \emph{p-n} interfaces.

Another interesting situation for four-terminal case is that when a small bias voltage $V_0$ is applied to the terminals 2 and 4 instead of 1 and 3, the QH resistance $R_{24,13}$ and the longitudinal resistance $R_{24,24}$ can be measured. It is found that the QH resistances in the two cases are opposite to each other $R_{13,24}=-R_{24,13}$ but the longitudinal resistances $R_{13,13}$ and $R_{24,24}$ are completely different.

Now we explore the six-terminal Hall device [see Fig.~\ref{fig1}(b)], where a small basis $V_0$ is
applied to the longitudinal terminals 1 and 4. Here the QH and longitudinal resistances are defined by $R_{14,26}=(V_2-V_6)/I_1$ and $R_{14,23}=(V_2-V_3)/I_1$. The resistances satisfy the properties $R_{14,26}=R_{14,35}$ and
$R_{14,23}=R_{14,65}$. Similarly, there are nine different regimes and we still explore the resistances $R_{14,26}$ and
$R_{14,23}$ in the six regimes for $V>0$, shown as below. A straightforward calculation gives,

(a) ee-ee-ee regime ($|\nu_{1\uparrow}|\geq|\nu_{2\uparrow}|$, $|\nu_{1\downarrow}|\geq|\nu_{2\downarrow}|$):
\begin{eqnarray}
&&R_{14,26}=\frac{h}{e^2}\frac{1}{|\overline{\nu_1}|},\label{eq:17}\\
&&R_{14,23}=\frac{h}{e^2}(\frac{1}{|\overline{\nu_2}|}-\frac{1}{|\overline{\nu_1}|});\label{eq:18}
\end{eqnarray}
(b) eh-eh-eh regime ($|\nu_{1\uparrow}|\geq|\nu_{2\uparrow}|$, $|\nu_{1\downarrow}|\leq|\nu_{2\downarrow}|$):
\begin{eqnarray}
&&R_{14,26}=\frac{h}{e^2}\frac{|\nu_{1\uparrow}||\nu_{2\uparrow}|-|\nu_{1\downarrow}||\nu_{2\downarrow}|}
{|\overline{\nu_1}||\nu_{2\downarrow}|(|\nu_{2\downarrow}|-|\nu_{2\uparrow}|)
+|\overline{\nu_2}||\nu_{1\uparrow}|^2},\label{eq:19}\\
&&R_{14,23}=\frac{h}{e^2}\frac{|\nu_{1\uparrow}|(|\nu_{1\uparrow}|+|\nu_{2\downarrow}|-|\nu_{2\uparrow}|)}
{|\overline{\nu_1}||\nu_{2\downarrow}|(|\nu_{2\downarrow}|-|\nu_{2\uparrow}|)
+|\overline{\nu_2}||\nu_{1\uparrow}|^2};\label{eq:20}
\end{eqnarray}
(c) ee-eh-ee regime ($|\nu_{1\uparrow}|\geq|\nu_{2\uparrow}|$):
\begin{eqnarray}
&&R_{14,26}=\frac{h}{e^2}\frac{|\nu_{2\uparrow}|}
{|\nu_{2\downarrow}|(|\nu_{2\downarrow}|-|\nu_{2\uparrow}|)+|\overline{\nu_1}||\overline{\nu_2}|},\label{eq:21}\\
&&R_{14,23}=\frac{h}{e^2}\frac{|\overline{\nu_1}|+|\nu_{2\downarrow}|-|\nu_{2\uparrow}|}
{|\nu_{2\downarrow}|(|\nu_{2\downarrow}|-|\nu_{2\uparrow}|)+|\overline{\nu_1}||\overline{\nu_2}|};\label{eq:22}
\end{eqnarray}
(d) hh-hh-hh regime ($|\nu_{1\uparrow}|\leq|\nu_{2\uparrow}|$, $|\nu_{1\downarrow}|\leq|\nu_{2\downarrow}|$):
\begin{eqnarray}
&&R_{14,26}=-\frac{h}{e^2}\frac{1}{|\overline{\nu_2}|},\label{eq:23}\\
&&R_{14,23}=0;\label{eq:24}
\end{eqnarray}
(e) eh-hh-eh regime ($|\nu_{1\downarrow}|\leq|\nu_{2\downarrow}|$):
\begin{eqnarray}
&&R_{14,26}=-\frac{h}{e^2}\frac{|\nu_{1\downarrow}|}
{|\nu_{1\uparrow}|^2+|\overline{\nu_1}||\overline{\nu_2}|},\label{eq:25}\\
&&R_{14,23}=\frac{h}{e^2}\frac{|\nu_{1\uparrow}|^2+|\nu_{1\uparrow}||\overline{\nu_2}|}
{(|\nu_{1\uparrow}|^2+|\overline{\nu_1}||\overline{\nu_2}|)|\overline{\nu_2}|};\label{eq:26}
\end{eqnarray}
(f) ee-hh-ee regime:
\begin{eqnarray}
&&R_{14,26}=0,\label{eq:27}\\
&&R_{14,23}=\frac{h}{e^2}\frac{1}{|\overline{\nu_2}|}.\label{eq:28}
\end{eqnarray}

\begin{figure}
\scalebox{0.82}[0.87]{\includegraphics[17,15][312,213]{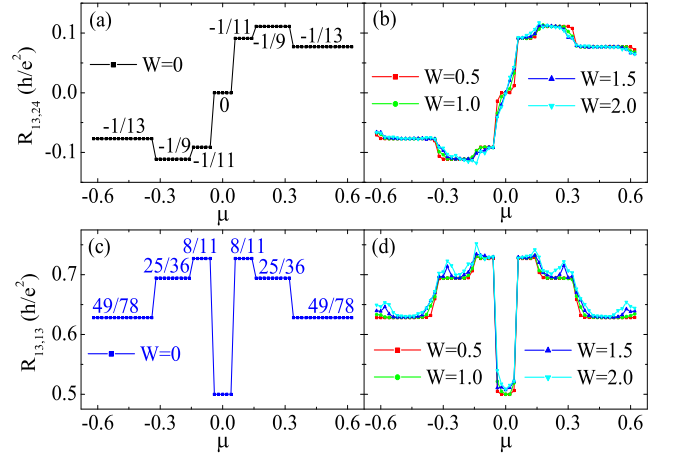}}  \caption{\label{fig5}(Color online) The numerical QH resistance $R_{13,24}$ [(a) and (b)] and longitudinal resistance $R_{13,13}$ [(c) and (d)] vs the chemical potential $\mu$ for different disorder strength $W$ at $M=0.6$, $eV=0.45$ and $\phi=0.01$. The size of the center block region are $9.7$~nm$\times10.7$~nm (40 armchair chains $\times$60 zigzag chains). The corresponding magnetic length is $l_B=\sqrt{\hbar /(eB)}=0.9$~nm which is much smaller than the size, so the edge states can propagate freely in the absence of disorder along the edges or interfaces without mixing with each other. }
\end{figure}

To make sure the topological stability of the predicted quantized resistance, it is necessary to make a numerical calculation to discuss the effect of disorder on the QH and longitudinal resistances.\cite{Chen2010,Fogler2008,Long2008} For the four-terminal case discussed here, we assume that the disorder only exists in the center block region [dashed-box region shown in Fig.~\ref{fig1}(a)].\cite{Chen2010,Sun2010,Long2008} In the presence of the disorder, the on-site energy is changed to $\epsilon_i+w_i$, where $w_i$ is the on-site disorder energy which is randomly distributed within the range $[-W/2,W/2]$ with $W$ the disorder strength. Since there exists no conducting channel between terminal 1 and 3, $T_{13}$=$T_{31}$=0. Generally, the QH and longitudinal resistances can be expressed as:
\begin{eqnarray}
R_{13,24}=\frac{T_{21}T_{43}-T_{23}T_{41}}{T_{12}(T_{24}T_{43}+T_{23}T_4)+T_{14}(T_{42}T_{23}+T_{43}T_2)},\\
R_{13,13}=\frac{T_2T_4-T_{24}T_{42}}{T_{12}(T_{24}T_{43}+T_{23}T_4)+T_{14}(T_{42}T_{23}+T_{43}T_2)},
\end{eqnarray}
where we introduced $T_4=T_{41}+T_{42}+T_{43}$ and $T_2=T_{21}+T_{23}+T_{24}$.
Here, $T_{pq}=\sum_\sigma \mathrm{Tr}[\mathbf{\Gamma}_{p\sigma}\mathbf{G}_\sigma^R\mathbf{\Gamma}_{q\sigma}\mathbf{G}_\sigma^A]$
($p,q=1,2,\cdots,4$ and $p\neq q$), with $\mathbf{\Gamma}_{p\sigma}=i[\mathbf{\Sigma}_{p\sigma}^R-\mathbf{\Sigma}_{p\sigma}^A]$, the retarded and advanced Green's functions $\mathbf{G}^R=[\mathbf{G}^A]^\dag=[E-\mathbf{H}_\sigma^c-\sum_p\mathbf{\Sigma}_{p\sigma}^R]^{-1}$. $\mathbf{H}_\sigma^c$ is the Hamiltonian for the center block region. The retarded self-energy $\mathbf{\Sigma}_{p\sigma}^R$ due to the coupling to terminal $p$ can be calculated numerically by the recutisive Green's function technique.\cite{Lee1981} Figures \ref{fig5}(a) and \ref{fig5}(c) show numerical results for the resistances in the absence of disorder $W=0$. Compared with the above analytic result, these numerical results give exactly the same plateau values [see the plateau values of one-dimensional slices in Fig.~\ref{fig4}(e)]. The figures also illustrate the antisymmetric and symmetric features of $R_{13,24}$ and $R_{13,13}$ with respect to $\mu=0$ respectively. In Figs.~\ref{fig5}(b) and \ref{fig5}(d), we show the resistances for different disorder strengths $W$. The plateau structures of $R_{13,24}$ and $R_{13,13}$ still exist for weak and moderate disorder, and the quantum plateau values are the same as that of the clean graphene \emph{p-n} junction. This indicates that the quantum plateaus of $R_{13,24}$ and $R_{13,13}$ are quite stable and robust against disorder due to the topological stability of the system. However, for a very strong disorder (e.g. $W\geq2$), the plateau structures of $R_{13,24}$ and $R_{13,13}$ will finally be smeared out and then disappear.

In summary, we have studied the QH and longitudinal resistances in the four- and six-terminal FGPNJs under a perpendicular magnetic field. In the proposed Hall devices, the local filling factors are independently controlled by the gate voltages and the transverse contacts are arranged at \emph{p-n} interface. Owing to the competition between spin splitting and local potentials, many new quantum plateaus of resistances are revealed. By avoiding the edge states mixing at the interface, the resistances are topologically protected and so robust against weak disorder. This multi-terminal transport study on the FGPNJ may provide a deeper insight into the transport properties of the FGPNJ, and the newly predicted fractional resistances are expected to be observed in experiment.

\begin{acknowledgments}
The author L. Xu thanks F. Lu for useful discussions. This work was supported by NSFC Project Nos. 10504009 and 10874073, and 973 Project Nos.
2009CB929504, and 2011CB922101.
\end{acknowledgments}


\end{document}